\newcommand{\tag}[1]{[{\it #1}]}
\newcommand{\ol}{\overline}
\newcommand{\kz}{Knizhnik-Zamolodchikov \relax}

\documentstyle [12pt]{article}
\newlength{\extraspace}
\newlength{\extraspaces}
\newcommand{\beq}{\begin{equation}
\addtolength{\abovedisplayskip}{\extraspaces}
\addtolength{\belowdisplayskip}{\extraspaces}
\addtolength{\abovedisplayshortskip}{\extraspace}
\addtolength{\belowdisplayshortskip}{\extraspace}}
\newcommand{\eeq}{\end{equation}}

\newcommand{\beqa}{\begin{eqnarray}
\addtolength{\abovedisplayskip}{\extraspaces}
\addtolength{\belowdisplayskip}{\extraspaces}
\addtolength{\abovedisplayshortskip}{\extraspace}
\addtolength{\belowdisplayshortskip}{\extraspace}}
\newcommand{\eeqa}{\end{eqnarray}}

\newcommand {\dell}[1]{\frac{\partial}{\partial #1}}

\newcommand{\fpar}{\hspace{6mm}}

\newcommand{\ie}{{\it i.e}.\ }

\newcommand{\is}{\! & \! = \! & \!}
\newcommand{\n}{\nonumber \\[1.5mm]}
\newcommand{\half}{{1 \over{2}}}

\newcommand{\definition}[2]{
\begin{figure}[htb]
\vspace{#1mm}
\begin{center}
\setlength{\unitlength}{0.6mm}
{\mbox{\begin{picture}(102,100)(0,0)
\thinlines
\put(-40, 30){\vector(0,1){10}}
\put(-40, 30){\vector(1,0){12}}
\put(-40, 30){\vector(2,3){4}}
\put(-38, 40){\makebox(0,0){$t$}}
\put(-26, 30){\makebox(0,0){$x$}}
\put(-32 ,36){\makebox(0,0){$y$}}
\put(-9, 30){\line(1,0){120}}
\put(-9, 90){\line(1,0){120}}
\put(115, 28){\makebox(0,0)[bl]{$t_0$}}
\put(115, 88){\makebox(0,0)[bl]{$t_1$}}
\thicklines
\put(1,30){\line(0,1){60}}
\put(16,30){\line(0,1){60}}
\put(46,30){\line(1,2){30}}
\put(61,30){\line(0,1){28}}
\put(61,62){\line(0,1){28}}
\put(101,30){\line(0,1){60}}
\put(31,60){\circle*{1}}
\put(41,60){\circle*{1}}
\put(71,60){\circle*{1}}
\put(81,60){\circle*{1}}
\put(91,60){\circle*{1}}
\put(0,20){\makebox(0,0)[bl]{1}}
\put(15,20){\makebox(0,0)[bl]{2}}
\put(45,20){\makebox(0,0)[bl]{$i$}}
\put(58,20){\makebox(0,0)[bl]{$i+1$}}
\put(100,20){\makebox(0,0)[bl]{$N$}}
\parbox{16cm}{\small #2}
\end{picture}}}
\end{center}
\end{figure}}

\begin{document}

\begin{flushright}
{\sc NUS/HEP/}920502\\
May 1992
\end{flushright}
\vspace{.3cm}

\begin{center}
{\Large{\bf{Mutual statistics, braid group, and \\[2mm]
the fractional quantum Hall effect}}}\\[13mm]

{\sc Christopher Ting}\footnote{e-mail: scip1005@nuscc.nus.sg}\footnote{
Part-time affiliation:
Department of Physics, National University of Singapore}\\[5mm]
{\it Defence Science Organization, \\[2mm]
20 Science Park Drive, Singapore 0511 \\[2mm]
The Republic of Singapore } \\[35mm]
{\bf Abstract}
\end{center}
\noindent
We show that the notion of mutual statistics arises naturally
from the representation theory of the braid group over the multi-sheeted
surface. A Hamiltonian which describes particles moving on the
double-sheeted surface is proposed as a model for
the bilayered fractional quantum Hall effect (FQHE) discovered recently.
We explicitly show that the quasi-holes of the bilayered Hall fluid
display fractional mutual statistics. A model for
3-dimensional FQHE using the multi-layered sample is suggested.
\vfil
\newpage
\section{Introduction}
\fpar
Anyons, or particles with quantum statistics interpolating between
bosons and fermions \\ \cite{Wilczek},
are associated with interesting 2-dimensional
condensed matter physics. By now,
it is quite established that the quasi-particle and quasi-hole
excitations of the fractional quantum Hall effect (FQHE) obey
anyonic statistics. The experiments probing the behaviour of the
longitudinal resistance \cite{long} , together with those studying the
transmission probability and reflection probability between two
electron liquids \cite{Chang-Cunningham} provide clear evidence that
the quasi-excitation is fractionally charged,
a fact best described by a theory which treats the quasi-excitation
as anyon \cite{Laughlin}.
The studies of anyons have also led to a novel model of
superconductivity \cite{superconductivity,supcond},
which may or may not have any relevance to the physics of
superconducting materials with high critical temperature.
It is also noteworthy that the concept of anyon or
braid group statistics has been generalized to the
nonabelian case \cite{nonabelian,Lai-Ting},
and particles on Riemann surfaces of higher genus \cite{higher genus}.

Though the possibility of such exotic statistics is peculiar to
2-dimensional many-body quantum system,
there is still one aspect of quantum statistics that has not been well
excogitated.
One is used to the notion that quantum statistics is about particles
being indistinguishable. However, in two dimensional systems, even
particles which are distinguishable can be assigned with a ``statistics"
when they exchange their positions.

Recently,
with the experimental discoveries of FQHE at filling factor $\nu = \half$,
Wilczek went further and introduced the term ``mutual statistics"
\cite{Wilczek-92,genCS}.
It is a novel idea prompted by the observation of this new fractional
quantum Hall state in {\em double}-layered two-dimensional electronic systems.

The experimental realization of a $\nu = \half$ FQHE has been much sought
after and its discovery is exciting indeed.
Two research groups, Suen {\it et al.} \cite{Suen} and
Eisenstein {\it et al.} \cite{Eisenstein}
have independently found the $\nu = \half$
magneto-transport spectrum, albeit with different sample geometries. While
on the scale of the magnetic length $\ell$,
the former group used a wide GaAs/AlGaAs single quantum well to
realize the bilayered electronic system,
the latter's samples contained two narrow quantum wells separated by
an undoped pure AlAs barrier layer. At this stage,
it is not clear whether these
two states with the same filling factor are in the same universality
class \cite{GWW}. Certainly, different geometries give rise to
different profiles of the distribution functions of the system along
the direction normal to the surface of the sample. Nevertheless, for
the discussion of mutual statistics, we shall for the time being
assume that the electronic system is ideally bilayered.

The crucial thing about the bilayered system is
that one can have two kinds of electrons, by introducing
an index to indicate which of the two layers the electrons reside.
Furthermore, the layer index can be treated as a kind of quantum number,
called ``pseudo-spin", in close
analogy to the spin of the electrons\footnote{It
is justified to ignore the spin degree of freedom as all the electrons
are polarized by the strong magnetic field.}. Thus, the bilayered
electronic system possesses an additional degree of freedom which, as
discussed by Yoshioka, MacDonald and Girvin (YMG) \cite{YMG},
allows Laughlin wavefunction to abide in Pauli principle yet describes a
FQHE state with {\em even} denominator filling factor.

Suppose we label the coordinates of those electrons in the first layer with
$w_a^{\tag{1}}$ and those in the second layer $w_a^{\tag{2}}$.
The ans\"atz $\psi_{m_1, m_2, n}$
proposed by YMG is an almost trivial generalization of
the Laughlin wavefunction:
\beqa
\psi_{m_1, m_2, n} ( w_a^{\tag{1}}, \, w_a^{\tag{2}} )
= \! \! & & \! \! \prod_{a < b} (w_a^{\tag{1}} - w_b^{\tag{1}} )^{m_1}
\prod_{a < b} (w_a^{\tag{2}} - w_a^{\tag{2}} )^{m_2}
\prod_{a, \, b} ( w_a^{\tag{1}} - w_b^{\tag{2}} )^n \\
\!\! & & \!\! \times \,
\exp  - \frac{1}{4 \ell^2} \left( \sum_a^{N^{\tag{1}}} |w_a^{\tag{1}}|^2
+ \sum_a^{N^{\tag{2}}} |w_a^{\tag{2}}|^2  \right) \, .
\label{YMG}
\eeqa
The integers $m_1$ and $m_2$ are odd, as entailed by the
fact that the electrons in the same layer are indistinguishable.
The non-trivial aspect of $\psi_{m_1, m_2, n}$ as a generalization
of Laughlin wavefunction lies in the factor
$\prod_{a, \, b} ( w_a^{\tag{1}} - w_b^{\tag{2}} )^n$
which describes some sort of
inter-layer correlation. As their numerical studies indicate,
the overlap between the ground state wavefunction of a few
electrons interacting via Coulomb potential
and $\psi_{m_1, m_2, n}$ is dependent on
$\delta = \frac{d}{\ell}$, where $d$ denotes the distance between the layers.

It this article, we shall discuss the topological order of the bilayered
FQHE, using the braid group approach we did for single-layered
FQHE states \cite{Ting-Lai-1,Ting-Lai-2}. We are interested to
find out if the inter-layer correlation characterized by
$\prod ( w_a^{\tag{1}} - w_b^{\tag{2}} )^{n}$
can be understood within the more fundamental
formulation of anyon physics, namely, the theory of the
quantum statistics in two spatial dimensions as the representation
theory of braid group \cite{Wu}. The key idea involves generalizing the path
integral representation of Artin's braid group to an analogue associated
with many-sheeted surfaces. As we shall see,
the idea of particles moving freely in the double-sheeted surface
gives rise to the notion of mutual statistics.

The paper is presented as follows. In the next section, we review the
path integral representation of the braid group \cite{Lai-Ting}
and its application \cite{Ting-Lai-1} to the FQHE of Laughlin type.
Using the same method, we proceed to construct a similar representation for
the braid group of the double-sheeted plane in section 3. The motivation
is to capture the physics of bilayered Hall media within the framework of
braid group statistics. In
section 4, we show that $\psi_{m_1, m_2, n}$ is an exact solution
of the ground state equations. Explicitly, we also illustrate
how mutual statistics arises in the braid group approach.
In section 5, we discuss various aspects of
the multi-layered FQHE. Our method allows us to calculate the
filling factors of samples which have arbitrary number of layers. We also
touch on the notion of 3-dimensional FQHE. Section 6 summarizes the main
results of this paper.

\section{Puncture phase and the braid group}
\fpar
The significance of the braid group for quantum statistics of particles
living in two spatial dimensions is clearly discussed by Y. S. Wu
\cite{Wu} who constructed a one-dimensional (abelian)
representation in the path integral formalism. The braid group
$B_N$ of $N$ threads are generated by
generators $\sigma_i$, $i = 1, \cdots, N - 1$ which are defined as
the $i$-th thread crosses {\em over} the $i+1$-th thread (Fig. 1).
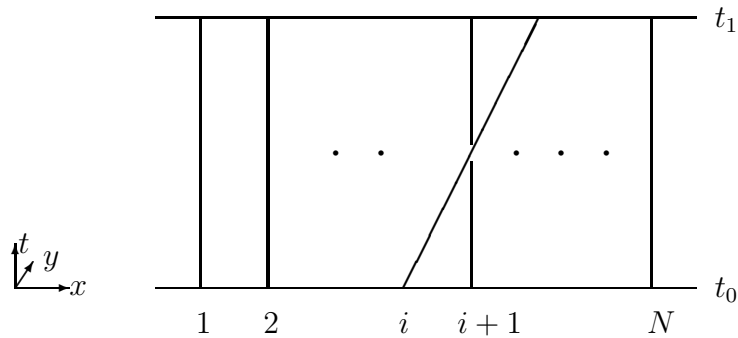
\definition{0}{\hspace{10mm} Fig. 1: The definition of $\sigma_i$.}

They satisfy the following algebraic relations:
\beq
        \sigma_{i} \sigma_{j} = \sigma_{j} \sigma_{i}, \hspace{20mm}
           |i - j| \geq 2,
        \label{ijji}
\eeq
\beq
        \sigma_{i} \sigma_{i+1} \sigma_{i}
                  =  \sigma_{i+1} \sigma_{i} \sigma_{i+1}, \hspace{10mm}
           i = 1, \cdots, n - 2.
        \label{ad}
\eeq
The basic idea of the path integral representation \cite{Wu} is to see
the threads as non-relativistic world lines of point particles.
Equally important is the observation that the underlying topology
of the 2-dimensional plane is non-simply connected when
each particle sees others as punctures \cite{Lai-Ting}.

\subsection{Non-abelian representation of the braid group}
\fpar
Because the configuration
space has become multiply connected, the paths are homotopically
classified and those of different classes cannot be smoothly
deformed from one to the other. When one considers the
Feynman kernel for a particle moving from point $z_{a(0)}$
at time $t_0$ to point $z_{a(1)}$ at time $t_1$,
one has to organize the paths according to their homotopical
classes. Now, the homotopy class of a path in the $N-1$-punctured
configuration space $M_{N-1}$
is determined by the winding numbers with respect to the
punctures.

It must be emphasized that
the notion of winding around a marked point (puncture)
is valid only if the space is
2-dimensional .
We formulate the homotopic constraint by introducing the notion of
{\em charged} winding ``number":
\beq
      w = \frac{1}{2\pi i} \int_{C} \frac{dz_a}{z_a - z_b}
                                  T_a \otimes T_b.
\label{windnum}
\eeq
Here,
$T_a$ and $T_b$ are the representations carried by the particles.
We see the threads as more than merely worldlines.
When the particles carry non-abelian charges, they become Wilson lines.
The corresponding charged winding angle $\theta$ can now be defined as:
\beq
       \theta = \hbox{sign}(C)~|\theta_{a(1)} - \theta_{a(0)}| + 2\pi w.
       \label{wind}
\eeq
We choose the convention that a path going counterclockwise about
the puncture
$z_b$ has positive sign, namely $\hbox{sign}(C) = 1$, and
denote $\vartheta = \hbox{sign}(C)~|\theta_{a(1)} - \theta_{a(0)}|$.
With this convention,
for the homotopically equivalent paths corresponding to
$\sigma_i$, which cross over from the left, the
change in the azimuthal angle $\vartheta$ is non-negative.
Compared to Wu's \cite{Wu} original construction,
Our $\vartheta$ is fixed by the initial and final positions
of the particle and does not play a significant role here.
The constrained Feynman kernel of homotopy class $l$ for particle
$a$ with mass $m$ can be expressed formally as:
\beq
     K_{l}(z_{a(1)}, t_{1}, z_{a(0)}, t_{0}) =
       \int {\cal D}_{l}z_a(t) {\cal D}_{l}\overline{z}_a(t)
       \exp~i\int_{t_{0}}^{t_{1}}\frac{1}{2} m |\dot{z}_a(t)|^2 dt~~
       \delta^2 ( 2\pi l~T_{a} \otimes T_{b}- \theta ).
       \label{Fk}
\eeq
With the path ordering determined by that in the definition
of charged winding angle
(\ref{wind}), the matrix-valued Dirac delta function can be
represented by the following path-ordered Fourier transform:
\beq
   \delta^2 ( 2\pi l~T_{a} \otimes T_{b}- \theta ) = \int\!\!\int
                \frac{dk}{2\pi} \frac{d\overline{k}}{2\pi}\,
                e^{-i(k\vartheta + {\overline k}\vartheta )}
          \,\hbox{P} \exp~i\left[ 2\pi k(l~T_{a} \otimes T_{b} - w)
                + a.c.\right].
          \label{delta_function}
\eeq
We use $a.c.$ to denote the anti-chiral copy of the preceding terms within
the delimiters.

Substituting (\ref{delta_function})
into the Feynman kernel (\ref{Fk}), we obtain the Fourier
transform:
\beq
     K_{l}(z_{a(1)}, t_{1}, z_{a(0)}, t_{0}) =
        \int\!\!\int \frac{dk}{2\pi} \frac{d\overline{k}}{2\pi}~
                e^{-i(k\vartheta + \overline{k}\vartheta )}\,
                \widetilde{K}_{l}(z_{a(1)}, t_{1}, z_{a(0)},
                 t_{0}; k, {\overline k}),
    \label{Ft}
\eeq
where
\beqa
   \widetilde{K}_{l}(z_{a(1)},t_{1},z_{a(0)},t_{0};k,\overline{k})
   \!\! &=& \!\!
   \int {\cal D}_{l}z_a(t) {\cal D}_{l}\overline{z}_a(t)\,
   \hbox{P}\! \exp \, i\int_{t_{0}}^{t_{1}}
   \frac{1}{2} m |\dot{z}_a(t)|^2~dt \n
   \!\! & & \!\! \times\, \exp \, i \, \int_{t_{0}}^{t_{1}} \left(
   k( \frac{ i \, \dot{z}_a}{z_a - z_b} + 2\pi  l )
   T_{a} \otimes T_{b} + a.c. \right)~dt.\n
   \label{tildeFk}
\eeqa

Expressions (\ref{Fk}) and (\ref{tildeFk}) can be easily generalized
to $N$ particles at $z_1, z_2, \cdots , z_N$, with
$\mbox{\rm Re} \, z_1 < \mbox{\rm Re} \, z_2
< \cdots < \mbox{\rm Re} \, z_N$.
Let particle $i$ make a trip from $z_{i \, (0)} = z_{i}(t_{0})$ to
$z_{i \, (1)} = z_{i}(t_{1}), \, \mbox{\rm Re}
\, z_{i \, (1)} > \mbox{\rm Re} \, z_{i+1}.$
Denoting the difference in the initial angle and the
final angle of the paths of particle $i$ with respect to particle $j$ as
$\vartheta_{ij}$,
$\vartheta_{ij} =\hbox{sign}(C_{i})~|\theta_{ij \, (1)} -
\theta_{ij \, (0)}|$,
the constrained Feynman kernel of
homotopy class $( l_{1},\cdot\cdot, l_{i-1}, l_{i+1}, \cdot\cdot, l_{n})$ for
particle $i$ carrying representation $T_i$ is
\beqa
    & & K_{l_{i}}(z_{i \, (1)}, t_{1}, z_{i \, (0)}, t_{0}) = \n
    & &\qquad\int\!\!\int \frac{dk}{2\pi} \frac{d\overline{k}}{2\pi}~
                \exp \left(-i \sum_{j =1, j \ne i}^n
                       \left( k\vartheta_{ij} + \overline{k}\vartheta_{ij}
                       \right) \right)
       \widetilde{K}_{l_{i}}(z_{i \, (1)}, t_{1}, z_{i \, (0)}, t_{0}; k,
       \overline{k}),
\eeqa
where
\beqa
     \widetilde{K}_{l_{i}}(z_{i\, (1)},t_{1},
     z_{i\, (0)},t_{0};k,\overline{k}) &=&
     \int {\cal D}_{l_{i}}z_{i}(t) {\cal D}_{l_{i}}\overline{z}_{i}(t)\,
     \hbox{P}\,\exp~i\int_{t_{0}}^{t_{1}}
     \frac{1}{2} m_{i} |\dot{z}_{i}(t)|^2~dt  \n
     &\quad &\times \exp \, i \, \int_{t_{0}}^{t_{1}} \left(
     k \sum_{j = 1, j \ne i}^n
     \left( \frac{ i \, \dot{z}_{i}}{z_{i} - z_{j}} + 2\pi l_{j} \right)
     T_{i} \otimes T_{j} + a.c. \right) \, dt. \n
\eeqa
Given these initial and final conditions, $\sigma_{i}$ can be represented by
the positively oriented Feynman kernel of class
$(0, \cdot\cdot,\widehat{0}_{i}, \cdot\cdot, 0),$ the $i$-th 0
is omitted as we do not consider self-linking. Writing,
\beqa
                A_{z_{i}} \is i k \sum_{j=1, j \ne i}^N
                   \frac{ T_{i} \otimes T_{j}}{ z_{i} - z_{j}}, \\
                A_{\overline{z}_{i}} \is
                   i \overline{k} \sum_{j=1, j \ne i}^N
                   \frac{ \overline{T}_{i} \otimes \overline{T}_{j}}
                        { \overline{z}_{i} - \overline{z}_{j} },
\eeqa
the proposed representation $D(\sigma_{i})$ is
$K_{i}(t_{1},t_{0})$ given below;
\beqa
       \int {\cal D}_{+}z_{i} {\cal D}_{+}\overline{z}_{i}
       \int\!\!\int \frac{dk}{2\pi} \frac{d\overline{k}}{2\pi}~\hbox{P}\,
       \exp~i \, \int_{C_{i}}
       \frac{1}{2} m_{i} |dz_{i}|^2
       + A_{z_{i}} dz_{i} + A_{\overline{z}_{i}} d\overline{z}_{i} \n
       \times\,\exp \left(-i \sum_{j =1, j \ne i}^N
       \left( k\vartheta_{ij} + \overline{k}\vartheta_{ij} \right)
       \right),
       \label{rep}
\eeqa
followed by an exchange operation $\Pi_{i \, i+1}$,
\beq
      D(\sigma_{i}) = \Pi_{i \, i+1} \, K_i(t_1,t_0) \, .
\eeq
$\Pi_{i \, i+1}$ is to make every
world line stick to the same representation space it has
started with.
The multiplication rule for the braid group generators is realised as the usual
multiplication of kernels. One can prove that
$D (\sigma_i)$ satisfy the defining relations
(\ref{ijji}, \ref{ad}) \cite{Lai-Ting}.

Now, the effective Lagrangian of particle $i$ can be readily read from
(\ref{rep}).
\beq
    L = \frac{1}{2} m_{i} |\dot{z_{i}}|^2
        + A_{z_{i}} \dot{z}_{i} + A_{\overline{z}_{i}} \dot{\overline{z}}_{i}.
    \label{Lagrangian}
\eeq

The Schr\"{o}dinger equation associated to the Feynman kernel of particle
$i$ is
\beq
   i \frac{\partial}{\partial t} \psi = -\frac{1}{m_{i}}
   \left[ ( \partial_{z_{i}} - i A_{z_{i}} )
          ( \partial_{\overline{z}_{i}} - i A_{\overline{z}_{i}}) +
          ( \partial_{\overline{z}_{i}} - i A_{\overline{z}_{i}})
          ( \partial_{z_{i}} - i A_{z_{i}} ) \right] \psi.
  \label{Sch}
\eeq
In the limit $m_{i}\to 0,$ a
class of solutions of (\ref{Sch}) consists of those wavefunctions
$\psi$ satisfying
\beq
         ( \partial_{z_{i}} - i A_{z_{i}} )  \psi
         =  \left( \frac{\partial}{\partial z_{i}}
              + k \sum_{j=1, j \ne i}^N
                   \frac{ T_{i} \otimes T_{j}}{ z_{i} - z_{j}} \right) \psi
          =0 \, ,
         \label{KZ1}
\eeq
\beq
          ( \partial_{\overline{z}_{i}} - i A_{\overline{z}_{i}}) \psi
         = \left( \frac{\partial}{\partial \overline{z}_{i}}
             + \overline{k} \sum_{j=1, j \ne i}^N
                   \frac{ \overline{T}_{i} \otimes \overline{T}_{j}}
                        { \overline{z}_{i} - \overline{z}_{j} } \right) \psi
         = 0 \, .
        \label{KZ2}
\eeq
These are precisely the Knizhnik-Zamolodchikov equations if we set
$k = \overline{k}=-2/(l+c_V)$, where $l$ is the level of the WZW model
and $c_{V}$ is the quadratic Casimir of the adjoint representation
of the group $G$ \cite{KZ}. It is interesting to note that the
wavefunctions, though non-normalizable, are the parallel transport sections
of a complex vector bundle over the base manifold $M_{N-1}$.

In the context of particle
statistics, (\ref{Sch}) can be interpreted as the Schr\"odinger equation for
non-Abelian anyons. When $T_i = \overline{T}_i = 1$, $i = 1, \cdots, N$,
it is the (abelian)
1-dimensional irreducible representation constructed by Wu
\cite{Wu}. Therefore our construction is a non-Abelian generalization
of the general theory of quantum statistics in two dimensions.

\subsection{Laughlin ground state}
\fpar
The starting point of a plausible theory of FQHE is
Laughlin's ans\"atz \cite{Laughlin}:
\beq
|m \rangle
= \prod_{j<k} ( z_j - z_k )^m \exp (- \frac{1}{4 \ell^2} \sum_{i} | z_i |^2),
\label{L state}
\eeq
where $\ell = \sqrt{ \frac{ \hbar c }{e B}}$
is the magnetic length, $\hbar$, $c$ being the usual universal constants,
$e$ is the charge of the electron and $B$ is the strength of the magnetic
field. It is postulated that $|m \rangle$ is the ground state of the
electrons exhibiting FQHE with fractional filling factor $1 \over m$.
The reason why $m$ is odd is because $|m \rangle$ describes a system
of electrons which have fermionic statistics.

Let us consider the $N$-body Hamiltonian:
\beqa
\lefteqn{  H =
\frac{1}{m^{*}} \sum_{j=1}^N
[~( - i \hbar \partial_{z_{j}} + \frac{e}{c} B_{z_{j}} + A_{z_{j}} )
( - i \hbar \partial_{\overline{z}_{j}} + \frac{e}{c} B_{\overline{z}_{j}}
+ A_{\overline{z}_{j}}) } \n
& & \mbox{}
+ ( - i \hbar \partial_{\overline{z}_{j}} + \frac{e}{c} B_{\overline{z}_{j}}
+ A_{\overline{z}_{j}} )
( - i \hbar \partial_{z_{j}} + \frac{e}{c} B_{z_{j}}
+ A_{z_{j}} )~] \n
& =  &
\frac{\hbar^2}{m^{*}} \sum_{j=1}^N \left( D_{z_j} D_{\overline{z}_{j}} +
D_{\overline{z}_{j}} D_{z_j} \right) \, ,
\label{Ham}
\eeqa
where
\beqa
D_{z_j} & \equiv &
\partial_{z_j} + i \frac{e}{\hbar c} B_{z_j}
+ \frac{i}{\hbar} A_{z_j} \, , \\
D_{\overline{z}_{j}} & \equiv &
- \partial_{\overline{z}_{j}}
- i \frac{e}{\hbar c} B_{\overline{z}_{j}}
- \frac{i}{\hbar} A_{\overline{z}_{j}} \, ,
\eeqa
\beqa
A_{z_{j}} \is i m \hbar \sum_{k=1, k \ne j}^N
\frac{ T_{j} \otimes T_{k}}{ z_{j} - z_{k}}, \n
A_{\overline{z}_{j}} \is i \overline{m} \hbar \sum_{k=1, k \ne j}^N
\frac{ \overline{T}_{j} \otimes \overline{T}_{k}}{{\overline{z}_{j}} -
{\overline{z}_{k}}} \, ,
\eeqa
$m^*$ is the effective mass of the electron, and
$B_{z_j}, B_{{\overline z}_j}$ the components of the gauge field of
the external magnetic field.
The gauge field components $A_{z_j}$, $A_{{\ol z}_j}$
reflect the topological properties of the configuration space.
Our construction which hinges on formulating the
winding number as the homotopy label for the paths therefore is valid
if the width of the quantum well is smaller than, or of the order of
$2 \ell$. A non-trivial consequence is that one should expect the
single-layered FQHE to disappear when the width becomes unduly large.

Since all the particles are indistinguishable, they
carry the same representation. Thus, for any two
particles $k$, $j$, we have $T_j  = T_k$ and
$T_j = \overline{T}_j$, $j = 1, \cdots, N$. One may use hermitian
matrices
to represent $T_j^\alpha$, $\alpha = 1, \cdots, \mbox{\rm dim} \, G$.
In the symmetric gauge,
\beqa
B_{z_{j}} \is - i \frac{B}{4} \overline{z}_{j}\, , \n
B_{\overline{z}_{j}} \is  i \frac{B}{4} z_{j}\, , \n
m \is - \overline{m}\, ,
\label{eq: symmetric gauge}
\eeqa
one calculates
the commutator of $D_{z_i}$ and $D_{\overline{z}_{j}}$:
\beq
\left[ D_{z_j}, D_{\overline{z}_{j}} \right] = \frac{eB}{2 \hbar c}
       + 2 \pi m
       \sum_{ k = 1, k \neq j}^N \, \delta^{(2)} ( z_j - z_k )
       T_j \otimes T_k.
\eeq
The term $\frac{eB}{2 \hbar c}$ is related to the
zero-point energy of a simple harmonic oscillator, whereas
the Dirac delta functions arise from the
2-dimensional Green function of the plane:
\beqa
\partial_{\overline{z}} \frac{1}{ z - w} \is - \pi \delta^{(2)}
( z - w )\, , \\
\partial_z \frac{1}{ {\overline z} - {\overline w}}
\is - \pi \delta^{(2)} ( z - w )\, .
\eeqa
With $\omega \equiv \frac{e B}{m^* c}$, we can rewrite (\ref{Ham}) as
\beq
H =  \frac{2 \hbar^2}{m^{*}}
\sum_j D_{\overline{z}_{j}} D_{z_j} + {N \over 2} \hbar \omega
+ \frac{ 2 \hbar^2}{m^{*}}
\pi m \sum_j \sum_{k=1, k \neq j}
\delta^{(2)} ( z_j - z_k ) T_j \otimes T_k.
\label{Ham-delta}
\eeq
Since this Hamiltonian is derived from the assumption that the
underlying configuration space is not simply connected,
the ground state of $H$ can be obtained by letting
$z_j \neq z_k$ for all $j$ and $k$, and then consider
the following first order equation for $j$-th electron:
\beq
D_{z_j} \psi_{0 \, j} = \left[ \partial_{z_j}
+ \frac{e B}{4 \hbar c} {\overline{z}_{j}}
- m \sum_{k=1, k \neq j}^N
\frac{T_j \otimes T_k}{z_j - z_k} \right]
\psi_{0 \, j}= 0.
\label{eq: GSE}
\eeq
Writing
\beq
\psi_{0 \, j} =
       \exp ( - \frac{1}{4 \ell^2} |z_j|^2 )
       f_j(z_1, \cdots, z_N ) \, ,
\eeq
equation (\ref{eq: GSE}) then becomes
\beq
\partial_{z_j} f_j(z_1, \cdots, z_N )
- m \sum_{k=1, k \neq j}^N \frac{T_j \otimes T_k}{z_j - z_k}
f_j(z_1, \cdots, z_N ) = 0.
\label{eq: chiral KZ}
\eeq
Thus, we see that {\em chiral}
Knizknik-Zamolodchikov equations are relevant in FQHE.
For $T_j = 1, \, j = 1, \cdots N$,
the holomorphic function satisfying (\ref{eq: chiral KZ}) is
\beq
f_{j}(z_1, \cdots, z_N) = \mbox{\rm const}
\prod_{k = 1, k \neq j} ( z_j - z_k )^{m} \,.
\eeq
For $m > 0$, $f_j$ vanishes whenever $z_j$ coincides with any other $z_k$.
In other words, particle $j$ is kept apart from the other
electrons. This solution is consistent with the repulsive delta-function
potential $\sum_{ k = 1, k \neq j} \delta^{(2)} ( z_j - z_k )$,
because for any $j$,
\beq
\int d z_j d \overline{z}_j
\left(
\sum_{ k = 1, k \neq j} \!
\delta^{(2)} ( z_j - z_k )
\right) \, |f_j|^2 = 0.
\eeq
Though $f_j$ is not normalizable, $\psi_{0 \, j}$ is, thanks to
the factor $\exp ( - \frac{1}{4 \ell^2} |z_j|^2 )$ contributed by
the strong magnetic field. Solving $D_{z_j} \psi_{0 \, j} = 0$ for
arbitrary $j$, we find that the solution is exactly
the Laughlin wavefunction $|m \rangle$.

\subsection{Topological excitations}
In \cite{Laughlin}, Laughlin gave an ans\"atz of the wavefunction which
is a 1-quasi-hole excitation of the ground state $|m \rangle$ (\ref{L state}):
\beq
\psi_m (u; z_1, \cdots , z_N) = \prod_{j = 1}^N ( z_j - u )
|m \rangle \, ,
\eeq
where $u$ is the position of the quasi-hole. The existence of the quasi-hole
excitation is demonstrated in the {\em gedanken} experiment.
An infinitesimally thin solenoid is pierced through the ground state
$|m \rangle$ at position $u$. Adiabatically, a flux quantum $\frac{hc}{e}$
is added; $|m \rangle$ evolves in such a way that it remains an
eigenstate of the changing Hamiltonian. After the flux tube is
completely installed, the resulting Hamiltonian is related
to the initial one by a (singular) gauge transformation. To get
back to the original Hamiltonian, the
flux tube is gauged away, leaving behind an excited state
$\psi_m (u; z_1, \cdots, z_N)$.

Motivated by this physical picture, we
consider the same Hamiltonian (\ref{Ham}) for the
1-quasi-hole excitation but with a gauge transformed $A_{z_j},
A_{\overline{z}_j}$:
\beqa
A_{z_{j}} & \rightarrow & i m \hbar \sum_{k=1, k \ne j}^N
\left(
\frac{1}{ z_{j} - z_{k}}
+ \frac{\frac{1}{m}}{ z_{j} - u}
\right) \, ,
\n
A_{\overline{z}_{j}} & \rightarrow & -i m \hbar \sum_{k=1, k \ne j}^N
\left(
\frac{1}{{\overline{z}_{j}} - {\overline{z}_{k}}}
+ \frac{\frac{1}{m}}{{\overline{z}_{j}} - {\overline{u}}}
\right) \, .
\eeqa
It can be easily verified that $\psi_m$ satisfy the ground state
equations, $j = 1, \cdots, N$:
\beq
\left[ \partial_{z_j}
+ \frac{e B}{4 \hbar c} {\overline{z}_{j}}
- m \sum_{k=1, k \neq j}^N \left(
\frac{1}{z_j - z_k} + \frac{\frac{1}{m}}{z_j - u}
\right) \right]
\psi_m = 0.
\eeq
Recall that $A_{z_j}, A_{\overline{z}_j}$ originate from the
{\em charged} winding number constraint of
the paths of particle $j$ in the multiply-connected
configuration space. Attaching an additional solenoid on $|m \rangle$
therefore results in a new configuration space.
In other words,
electron $j$ sees the quasi-hole as a puncture as well, but
this time with charge $q_h = {1 \over m}$.
The excitation is topological in nature.
When a quasi-hole develops, the configuration space is topologically changed;
an additional puncture has appeared.

The scenario of many quasi-holes is given by the Halperin ans\"atz
\cite{Halperin}:
\beq
\psi_m( u_1, \cdots, u_{N_h} )
= \prod_{1 \leq j<k \leq N_h}
( u_j - u_k )^{1 \over m}
\exp (- \frac{1}{4 m \ell^2} \sum_{i} | u_i |^2)
\prod_{j, k} ( u_j - z_k )
|m \rangle \, .
\eeq
If we write
\beq
|\frac{1}{m} \rangle =
\prod_{1 \leq j<k \leq N_h}
( u_j - u_k )^{1 \over m}
\exp (- \frac{1}{4 m \ell^2} \sum_{i} | u_i |^2) \, ,
\eeq
which is of the same form as Laughlin's ground state $| m \rangle$,
we find that
\beq
\psi_m = |\frac{1}{m} \rangle \, | m \rangle
\prod_{j, k} ( u_j - z_k ) \, .
\eeq
Written in this form, the physical content of a collection of
quasi-holes $|\frac{1}{m} \rangle $
is explicit. They are just ``electrons" of (representation) charge
$q_h = {1 \over m}$ in the ``puncture" phase! The quasi-holes are also
under the influence of the external magnetic field. The exponential
factor $\exp (- \frac{1}{4 m \ell^2} \sum_{i} | u_i |^2)$ is
required to make the wavefunction normalizable.
The Hamiltonian for two species of electrons labelled by $q=1$
and $q_h = {1 \over m}$ is
\beqa
H \! &=& \!  \frac{ 2 \hbar^2}{m^{*}}
            \sum_j^N D_{\overline{z}_{j}} D_{z_j}
            + {N \over 2} \hbar \omega
            + \frac{ 2 \hbar^2}{m^{*}}
            \pi m \sum_j \sum_{k=1, k \neq j}
            \delta^{(2)} ( z_j - z_k )  \n
\! & \quad &\!
+ \frac{ 2 \hbar^2}{m_h^*}
\sum_j^{N_h} d_{\overline{u}_{j}} d_{u_j}
+ { N_h \over 2 } \hbar \omega_h
+ \frac{ 2 \hbar^2}{m_h^*}
\pi \frac{1}{m} \sum_j \sum_{k=1, k \neq j}
            \delta^{(2)} ( u_j - u_k )  \n
\! & \quad &\!
+  2 \hbar^2 \pi \left( { 1\over m^*} + {1 \over m_h^*} \right)
\sum_j^{N_h} \sum_k^N \delta^{(2)} ( u_j - z_k ) \, .
\label{eq: qhole}
\eeqa
where
\beqa
D_{z_j} \is
\partial_{z_j}
+ \frac{e B}{4 \hbar c} {\overline{z}_{j}}
- m \sum_{k=1, k \neq j}^N
\frac{1 \times 1}{z_j - z_k}
- m \sum_{k=1}^{N_h}
\frac{ 1 \times \frac{1}{m}}{z_j - u_k} \n
d_{u_j} \is
\partial_{u_j}
+ \frac{\frac{e}{m} B}{4 \hbar c} {\overline{u}_{j}}
- m \sum_{k=1, k \neq j}^{N_h}
\frac{\frac{1}{m} \times \frac{1}{m}}{u_j - u_k}
- m \sum_{k=1}^{N}
\frac{\frac{1}{m} \times 1}{u_j - z_k} \, ,
\eeqa
with similar expressions for $D_{{\overline z}_j}$
and $d_{{\overline u}_j}$. We have denoted the ``mass" of a
quasi-hole as $m_h^*$, and $\omega_h = \frac{1}{m} \frac{e B}{ m_h^* c}$ is
the angular frequency of the cyclotron motion of the quasi-holes
\cite{Ting-Lai-2}.
It can be readily shown that $\psi_m ( u_1, \cdots , u_{N_h}\, ; \,
z_1, \cdots, z_N )$ is the {\em exact} ground state solution of
$H$ (\ref{eq: qhole}).

\section{Particles on double-sheeted surface}
\fpar
So far, we only considered non-relativistic particles moving on a plane.
It is possible to generalize the representation theory
to particles moving on a many-sheeted
surface. The motivation for doing this is to suggest an alternative to the
notion of pseudo-spin \cite{YMG} in the discussion of FQHE of bilayered
samples. As will be discussed subsequently, we find that it is more natural
to follow the geometrical route, especially when the electronic system
contains more than two layers.

We shall begin with a short revision of Riemann's original
proposal \cite{Ahlfors}. Suppose we have a many-to-one mapping from
a $z$-plane to $w$-plane such as $w = z^n$, where $n > 1$ is an integer.
In this example, $n$ distinct points in the $z$-plane is mapped to
the same point in the $w$-plane. In other words, distinct points in
the $z$-plane have the same coordinates in the $w$-plane. Now,
suppose physical observations can only be done in the $w$-plane.
In order to reflect the true origin of the underlying geometry, which is
the $z$-plane, one can assign a tag to the points. It is clear that
with respect to the mapping $w = z^n$, the $z$-plane can be divided
into sectors each bounded by two angles $\frac{k-1}{n} 2 \pi$ and
$ \frac{k}{n} 2 \pi $, $k = 1, \cdots , n$. There is a one-to-one
correspondence between each sector and the $w$-plane. So one can assign a
tag $k$ to each point in the $z$-plane, according to the sector each belongs.
Those from the same sector carries the same tag. We say that
points with tag $k$ lie in the $k$-th sheet. The sheets
can be joined together at a ``cut" along the positive axis of the $w$-plane,
which gives rise to an upper and a lower ``edge" for each sheet.
The lower edge of the first sheet is ``glued" to the upper edge of the
second sheet, and the lower edge of the second sheet to the upper edge of
the third, and so on. Finally, the lower edge of the $n$th sheet is
glued to the upper edge of the first sheet.
With these formal ``cutting" and ``gluing" procedures,
the $w$-plane becomes a $n$-sheeted surface.

We are now ready to consider the path integral of particles on the
$n$-sheeted $w$-plane. For a start, we take $n = 2$ to keep the
equations from becoming bombastic. In complete analogy to the representation
theory studied in the last section, we assume that the particles moving
in the $w$-plane are in the puncture phase. Since $w$-plane is double-sheeted,
the coordinates of the particles carry either tag {\tag{1}}, or {\tag{2}}.
This is tantamount to saying that we have two kinds of particles taking
part in the puncture phase.

Let the number of particles with tag {\tag{1}} be $N^{\tag{1}}$
and that of particles with tag {\tag{2}} $N^{\tag{2}}$.
With the assignment of tags to the particles, we can provide similar
distinguishing label for the {\em charged} winding number $s^{\tag{ij}}$:
\beq
s^{\tag{ij}} = \frac{1}{ 2 \pi i}
\int_{C} \frac{d z_a^{\tag{i}}}{ z_a^{\tag{i}} - z_b^{\tag{j}} }
T_a^{\tag{i}} \otimes T_b^{\tag{j}} \, .
\eeq
where $ i = 1, 2; \, j = 1, 2$ are tag numbers,
and $a$, $b$ names given to the particles.
Notice that we have $s^{\tag{ij}} = s^{\tag{ji}}$ if all the particles
carry the same representation, \ie $T_a^{\tag{i}} = T_b^{\tag{j}}$ for
all $i, j$, $a$ and $b$.
Now, with the introduction of new tags $\tag{i}$,
the charged winding number becomes
\beq
\left( \begin{array}{cc}
s^{\tag{11}} & s^{\tag{12}} \\
s^{\tag{21}} & s^{\tag{22}}
\end{array}  \right) \, .
\label{D-winding}
\eeq
Correspondingly, the homotopic constraint is
$\delta^2 ( 2 \pi L - \Theta )$,
where $L$ is a matrix with 4 entries \footnote{If the charges are nonabelian,
each entry itself is a matrix.}:
\beq
L = (L^{\tag{ij}}) =
\left( \begin{array}{cc}
l^{\tag{11}} \, T_a^{\tag{11}} \otimes T_b^{\tag{11}} &
l^{\tag{12}} \, T_a^{\tag{12}} \otimes T_b^{\tag{12}}  \\
\vspace{0.5mm} & \vspace{0.5mm}                        \\
l^{\tag{21}} \, T_a^{\tag{21}} \otimes T_b^{\tag{21}} &
l^{\tag{22}} \, T_a^{\tag{22}} \otimes T_b^{\tag{22}}
\end{array}  \right) \, ,
\eeq
and $\Theta$ is also a matrix $\Theta = ( \theta^{\tag{ij}} )$ with
4 entries, each entry being
\beqa
\theta^{\tag{ij}} \is {\rm sign} (C) | \theta_{t_1}^{\tag{ij}} -
\theta_{t_0}^{\tag{ij}} | + 2 \pi s^{\tag{ij}} \,  \n
\!\! & \equiv & \!\! \vartheta^{\tag{ij}} + 2 \pi s^{\tag{ij}} \, .
\eeqa
The path-ordered Fourier transform of the constraint is,
\beq
\frac{1}{4 \pi^2}
\int\!\!\int d k^{\tag{ij}} {\ol k}^{\tag{ij}} \,
e^{ -i ( k^{\tag{ij}} \vartheta^{\tag{ij}} +
{\ol k}^{\tag{ij}} \vartheta^{\tag{ij}} )}
\, {\rm P}
\exp i [ 2 \pi k^{\tag{ij}} (  L^{\tag{ij}} -  s^{\tag{ij}} )
+ a.c. ] \, ,
\eeq
where the Lagrange multipliers $k^{\tag{ij}}$ form a matrix
in correspondence to (\ref{D-winding}):
\beq
\left( \begin{array}{cc}
k^{\tag{11}} & k^{\tag{12}} \\
k^{\tag{21}} & k^{\tag{22}}
\end{array} \right) \, .
\label{k-mat}
\eeq

In this way, we obtain the partition function of the paths
$w_a^{\tag{1}}$ belonging to the homotopy class $L$:
\beqa
\int {\cal D} w_a^{\tag{1}} {\cal D} {\ol w}_a^{\tag{1}}
\!\! & & \!\!
{\rm P} \exp i \int_{t_0}^{t_1} \half
m_a^{\tag{1}} |\dot{w}_a^{\tag{1}}|^2 dt \n
\!\! & & \!\! \times \, \,
\exp i \int_{t_0}^{t_1}
k^{\tag{11}} \left(
\frac{ i {\dot w}_a^{\tag{1}} }{ w_a^{\tag{1}} - w_b^{\tag{1}} }
+ 2 \pi l^{\tag{11}} \right)
T_a^{\tag{1}} \otimes T_b^{\tag{1}} \n
\!\! & & \!\! \hspace{15mm} + \, \,
k^{\tag{12}} \sum_b
\left(
\frac{ i {\dot w}_a^{\tag{1}} }{ w_a^{\tag{1}} - w_b^{\tag{2}} }
+ 2 \pi l^{\tag{12}} \right)
T_a^{\tag{1}} \otimes T_b^{\tag{2}} \n
\!\! & & \!\! \hspace{15mm} + \, \, a.c. \, dt \, .
\eeqa
The partition functions of the paths of
$w_b^{\tag{1}}$, $w_a^{\tag{2}}$ and $w_b^{\tag{2}}$
belonging to the homotopy class $L$ are similar.
These expressions may look complicated but they are based on the same idea
discussed in the previous section. The only difference of course is that
the paths are tagged and we have to consider them separately although they
are on the same plane.

Having discussed the puncture phase of four particles
$w_a^{\tag{1}}$,
$w_b^{\tag{1}}$, $w_a^{\tag{2}}$ and $w_b^{\tag{2}}$,
we can go on to consider the
multi-particle case. As anticipated, the gauge potential is
\beq
A_{w_a^{\tag{i}}} =  i k^{\tag{ij}}
\sum_{ b \neq a \, \, {\rm if} \, \, {\it i = j} } \!
\frac{ T_a^{\tag{i}} \otimes T_b^{\tag{j}} }
{ w_a^{\tag{i}} - w_b^{\tag{j}} }  \, ,
\label{gauge_pot_1}
\eeq
\beq
{\ol A}_{{\ol w}_a^{\tag{i}}} =  i {\ol k}^{\tag{ij}}
\sum_{ b \neq a \, \, {\rm if} \, \, {\it i = j} } \!
\frac{ {\ol T}_a^{\tag{i}} \otimes {\ol T}_b^{\tag{j}} }
{ {\ol w}_a^{\tag{i}} - {\ol w}_b^{\tag{j}} }  \, .
\label{gauge_pot_2}
\eeq

The novel feature of particles moving on a double-sheeted surface is
that the coupling constant of the gauge field is now a matrix (\ref{k-mat}).

This is the most important consequence of having two different categories of
particles on the same plane. For particles with tag $\tag{1}$,
they not only see each other as puncture
through the coupling constant $k^{\tag{11}}$,
but also those with tag $\tag{2}$ through $k^{\tag{12}}$. Similarly,
the same can be said for particles with tag $\tag{2}$.
If one considers a special case where
$k^{\tag{11}} = k^{\tag{12}} = k^{\tag{21}} = k^{\tag{22}}$,
one reverts back to the situation of particles moving on a
single-sheeted plane, which was the subject we have studied in section 2.

The non-trivial result of the present approach is clearly visible when
$k^{\tag{11}} \neq k^{\tag{12}}$ and
$k^{\tag{12}} \neq k^{\tag{22}}$. It gives rise to the so-called
mutual statistics \cite{Wilczek-92}, which is realized between the
quasi-excitations of the electron fluid in layer 1 and those in layer 2
of a bilayered sample exhibiting FQHE. This will be further discussed in
the next section when we consider the fractional
statistics of the quasi-hole excitation of the bilayered FQHE.

The braid group associated with particles moving on a double-sheeted
surface is generated by $\sigma_a$, $a = 1, \cdots, N^{\tag{1}} +
N^{\tag{2}} - 1$. Algebraically, the generators satisfy the same defining
relations. At the intuitive level, we can imagine weaving two kinds
of threads distinguished by their colours; we have $N^{\tag{1}}$ threads
of one colour and $N^{\tag{2}}$ threads of another colour. Though there is
nothing new about these generators, the representation theory using
path integrals however contains a richer flavour. We see that the
concept of mutual statistics emerges naturally from the framework.
The path integral approach compliments the field theoretic model based on
a {\em generalized} Chern-Simons action \cite{genCS}, as well as
the heuristic arguments of attaching {\em fictitious} flux tubes
on the charge carriers \cite{Wilczek}.

Let us briefly comment on the link with generalized Chern-Simons
theory \cite{genCS} which is defined by the following Lagrangian:
\beq
{\cal L} = \frac{1}{4 \pi} \sum_{i, j} \epsilon^{\mu \nu \lambda}
\alpha_{\mu}^{\tag{i}} \lambda^{\tag{ij}} \partial_{\nu}
\alpha_{\lambda}^{\tag{j}} \, .
\eeq
For simplicity, we only consider the $U(1)$ case. Different kinds of
gauge fields $\alpha^{\tag{i}}$ are involved here. They are coupled to each
other via coupling constants $\lambda^{\tag{ij}}$.
In the Coulomb gauge, $a_0^{\tag{i}} = 0$,
the induced Gauss law for the generalized system is
\beq
f^{\tag{i}}_{w_a^{\tag{i}} \, {\ol w}_a^{\tag{i}} }
= - 2 \pi \!
\sum_{ b \neq a \, \, {\rm if} \, \, {\it i = j} } \!
\sum_{j = 1}^2
(\lambda^{-1})^{\tag{ij}} \,
\delta^2 ( w_a^{\tag{i}} - w_b^{\tag{j}} ) \, ,
\eeq
where $i = 1, 2$, and
$ f^{\tag{i}}_{w_a^{\tag{i}} \, {\ol w}_a^{\tag{i}} }$ are the components of
the field strength. Implicitly, we have assumed that the matrix
$\lambda = ( \lambda^{\tag{ij}} )$ is invertible. If we equate
$(\lambda^{-1})^{\tag{ij}} = k^{\tag{ij}}$, we find that
the gauge potential (\ref{gauge_pot_1}, \ref{gauge_pot_2})
arising from the topological effect satisfies the induced Gauss law.

The Hamiltonian suggested by the path integral representation of the
braid group is
\beq
H_a^{\tag{1}} = - \frac{1}{m_a^{\tag{1}}}
\left( D_a^{\tag{1}} {\ol D}_a^{\tag{1}} + {\ol D}_a^{\tag{1}}
D_a^{\tag{1}} \right)
\label{H-double}
\eeq
for a free particle named $a$ and of tag ${\tag{1}}$.
In the Schr\"odinger representation, the differential
operator $D_a^{\tag{1}}$ is,
\beq
\partial_{w_a^{\tag{1}}} - i A_{w_a^{\tag{1}}}
\equiv
\dell{w_a^{\tag{1}}} + k^{\tag{11}} \sum_{b \neq a}^{N^{\tag{1}}}
\frac{ T_a^{\tag{1}} \otimes T_b^{\tag{1}} }{ w_a^{\tag{1}} - w_b^{\tag{1}}}
+ k^{\tag{12}} \sum_{b}^{N^{\tag{2}}}
\frac{ T_a^{\tag{1}} \otimes T_b^{\tag{2}} }{ w_a^{\tag{1}} - w_b^{\tag{2}}}
\, .
\label{D-double}
\eeq
The expression for ${\ol D}_a^{\tag{1}}$ is the anti-chiral
analogue of (\ref{D-double}).
The Hamiltonian for a free particle of tag ${\tag{2}}$ is of the same form
as (\ref{H-double}, \ref{D-double}), with $ 1 \leftrightarrow 2$.

\section{FQHE in bilayered systems}
\fpar
The recent discovery of $\nu = \half$ FQHE state on a bilayered system
\cite{Suen,Eisenstein} gives
supporting evidence to the theoretical prediction based on the
numerical studies of the YMG ans\"atz $\psi_{m_1, m_2, n}$
(\ref{YMG}) \cite{YMG} with $m_1 = m_2 = 3$ and $n = 1$.
The interesting feature of the bilayered system is that
electrons in different layers are also strongly correlated.
Among other things, the inter-layer correlation seems to be dependent on the
distance $d$ between the centres of the quantum wells. In a typical setup
where FQHE occurs, the magnetic length $\ell$ is about 100 \AA \relax.
The samples used in \cite{Eisenstein} have two 180-\AA \relax wide
GaAs quantum wells separated by an undoped AlAs layer 31 \AA \relax wide which
acts as a potential barrier. The experimental results show that
$\nu = \half$ effect is strongest when $d \over \ell$ is 2.4 and
the effect weakens when $d \over \ell$ increases. The result agrees
with YMG's numerical calculation of the overlap between the ground-state
wavefunction of a total of 6 electrons and $\psi_{3, 3, 1}$.
The key idea behind the ans\"atz lies in the prescription of
a pseudo quantum number to distinguish electrons in different layers.
The wavefunction $\psi_{m_1, m_2, n}$ was first conceived by Halperin
\cite{Halperin} to describe spin-unpolarized single-layered FQHE.
By definition, Halperin wavefunction has $m_1 = m_2$.
Furthermore, because the spin of an electron is a good quantum number, the
singlet condition forces $n = m_1-1$ \cite{Ting-Lai-2}.
In the case of spin-polarized
double-layered FQHE discussed by YMG, the restriction no longer applies,
as the electron-electron interaction depends on the layer in which the
electrons reside. In particular, the $\psi_{3 \, , 3 \, , 1}$ wavefunction
is not the eigenstate of the pseudo-spin, which means that it is {\em not}
the ground state in the limit $d \rightarrow 0$. So, the important
consequence of YMG ans\"atz is that the $\nu = \half$
FQHE state of Laughlin type can only be observed in the double-layered system.

\subsection{Ground state equations of the bilayered FQHE}
\fpar
As reviewed in section 2, we have produced a Hamiltonian for which
Laughlin wavefunction is the {\em exact} ground state. Naturally,
one wants to do likewise for $\psi_{m_1, m_2, n}$. Having constructed
the path integral representation of the braid group of double-sheeted
surface, it is possible and in fact rather straightforward to think of a
Hamiltonian which yields $\psi_{m_1, m_2, n}$ as ground state.

First, we shall point out some unusual aspect of $\psi_{m_1, m_2, n}$.
Strictly speaking, when $d \neq 0$, there is no
a priori reason why $\prod ( w_a^{\tag{1}} - w_b^{\tag{2}} )^n$ should appear.
In fact, even when $w_a^{\tag{1}} = w_b^{\tag{2}}$, the two electrons from the
respective layers are still a distance $d$ apart.
This is not true for electrons moving in the same layer.
Though energetically
unfavourable, electrons in the same layer can come arbitrarily close
to one another as in the event of collision;
there is no physical barrier to enforce that they be at least
a distance $d$ apart. How are we going to incorporate the asymmetry
between inter-layer and intra-layer electron-electron interaction?

In our proposal to understand the FQHE as a manifestation of the puncture
phase \cite{Ting-Lai-2},
we have particularly emphasized the role played by the cyclotron
motion. The point is that in order for the electrons to see each other as
punctures, we need each electron whirl with high frequency to effectively
exclude the region enclosed by the motion from being accessible to
other electrons. When the background magnetic field is high enough
such that this is achieved, literally the cyclotron motion has created
puncture. The physical intuition compares favourably with the fact that
a threshold in the background field strength $B$ exists, below which plateaux
dissolve; the excitation energy vanishes when $B$ drops below certain critical
value. Given this picture, one realizes that the electrons in the same
layer are forbidden to come close to each other; not closer than $2 \ell$
when the transition to puncture phase occurs. Viewed in this light,
the asymmetry alluded to disappears.

But this is only half the story. Remember that the layers are separated in
the direction of the magnetic field. The inter-layer and intra-layer
interactions are strictly speaking, mutually orthogonal.
Since the wavefunction $\psi_{m_1, m_2, n}$ does not contain information
about the coordinates in the direction of the magnetic field,
it is only an approximation of the actual system. Nevertheless, there
are reasons to believe that $\psi_{m_1, m_2, n}$ is a good approximation
of the FQHE ground state at the large scale.
Compared to the size of the sample, both $d$ and $\ell$ are small. They
can be seen as some UV cutoff of an effective field theory which is topological
in nature. One of the important insights springing from the use of
Chern-Simons action as an effective theory is that the Hall conductivity
of the system is identifiable as the inverse of the so-called level $k$ of the
topological gauge theory \cite{Frohlich-Kerler}, which is defined on
a space-time with two spatial dimensions. Thus, the width of the quantum
well is ignored. As it turns out, despite the cavalier tone of the
approximation, it is alright to do so when we restrict the discussion to
robust quantities such as the filling factor, and the statistics and charges
of the quasi-excitations.

{}From the perspective of braid group, we have also captured the
topological aspect of the FQHE. As reviewed in section 2, it was shown that
Laughlin wavefunction is an exact solution of the ground state equation.
Given that the FQHE is inherently associated with some topological order,
it is therefore tempting to find out if the inter-layer correlation of the
bilayered system is also of similar nature.

To begin with, we project the two layers onto a common 2-dimensional
plane which takes care of the remaining asymmetry between the inter-layer and
intra-layer correlation.
In other words, we assign a tag to indicate which of the two
layers the electrons are originally from.
Effectively, tagging the particles renders
the 2-dimensional plane multi-sheeted.
The tag is the only information one has about the direction
parallel to the magnetic field.
For a bilayered system, we have two kinds of particles on the plane.
Suppose the electron gas enters the puncture phase.
The quantum mechanics of this system of electrons with effective mass
$m^*$ is then given by the following Hamiltonian:
\beq
H = - \frac{1}{m^*} \left( \sum_a^{N^{\tag{1}}}
( D_a^{\tag{1}} {\ol D}_a^{\tag{1}} +  {\ol D}_a^{\tag{1}} D_a^{\tag{1}} )
+  \sum_a^{N^{\tag{2}}}
( D_a^{\tag{1}} {\ol D}_a^{\tag{2}} +  {\ol D}_a^{\tag{1}} D_a^{\tag{2}} )
\right), \,
\label{bilayer H}
\eeq
where in the symmetric gauge, the differential operator $D_a^{\tag{1}}$
is
\beq
D_a^{\tag{1}} \equiv
\dell{w_a^{\tag{1}}} + \frac{e B}{4 \hbar c} {\ol w}_a^{\tag{1}}
- k^{\tag{11}} \sum_{b \neq a}^{N^{\tag{1}}}
\frac{ 1 \times 1 }{ w_a^{\tag{1}} - w_b^{\tag{1}}}
- k^{\tag{12}} \sum_{b}^{N^{\tag{2}}}
\frac{ 1 \times 1 }{ w_a^{\tag{1}} - w_b^{\tag{2}}}
\eeq
The other operators ${\ol D}_a^{\tag{1}}$,  $D_a^{\tag{2}}$ and
${\ol D}_a^{\tag{2}}$ are similarly defined.
We can solve the ground state equations
\beq
D_a^{\tag{i}} \psi = 0
\eeq
for every $a$ and $i$. It is readily verifiable that $\psi_{m_1, m_2, n}$
is a solution of the equations, with
\beqa
k^{\tag{11}} \is m_1 \, , \\
k^{\tag{12}} \is n   \, , \\
k^{\tag{22}} \is m_2 \, .
\eeqa

The filling fraction of $\psi_{m_1, m_2, n}$ is
\beq
\nu = \frac{m_1 + m_2 - 2n}{\Delta} \, ,
\label{filling}
\eeq
where $\Delta$ is the determinant of the matrix (\ref{k-mat}), namely
\beq
\Delta = \det \,
\left( \begin{array}{cc}
m_1 & n \\
n   & m_2
\end{array} \right) \, .
\label{mat-double}
\eeq
$\nu$ is obtained as follows.
The inverse matrix of (\ref{mat-double}) is
\beq
\left( \begin{array}{cc}
\frac{m_2}{\Delta} & \frac{-n}{\Delta} \\
\vspace{-2mm}     & \vspace{-2mm}    \\
\frac{-n}{\Delta}  & \frac{m_1}{\Delta}
\end{array} \right) \, ,
\label{inverse}
\eeq
and $\nu$ is defined as the sum of all the entries of the
inverse matrix (\ref{inverse}).
The prescription to yield the filling faction $\nu$ is a generalization of
the single-layered case where $\nu = \frac{1}{m}$, \ie $\nu$ is (the sum of
the entries of) the inverse (matrix) of $m$, regarded as a
$1 \times 1$ matrix $m$.
As discussed earlier, the identification $\nu = m^{-1}$
has been made from the field theoretic point of view \cite{Frohlich-Kerler}.
It also agrees with the braid group approach, since in the
\kz equations (\ref{eq: chiral KZ}),
the level of the corresponding current algebra is the inverse of $m$
(see section 2).

Our result (\ref{filling}) agrees with that based on the intuitive arguments
\cite{Wilczek-92}.
Let us redo Wilczek's calculation of $\nu$ for the layered Hall media
\cite{Wilczek-92} in the braid group approach.
When the representation charge carried by the puncture is $U(1)$, as
in the case of polarized electrons in FQHE, there is a kind of ``duality"
between $m$ and the charge. For example:
\beq
\partial_{z} - m \sum_i^N \frac{ 1 \times 1}{z - w_i}
= \partial_{z} - \sum_i^N \frac{ 1 \times m}{z - w_i} \, .
\eeq
In words, particle $z$ of (representation) charge 1 unit sensing
punctures $w_i$ of charge 1 unit with the coupling strength $m$
is the same as sensing punctures of charge $m$ units with unit strength.
The total flux that particle $z$ sees is $m N$.
Using this ``duality", electrons of layer 1 sensing electrons of layer 2
with strength $n$ is equivalent to sensing ``fractionally charged"
electrons of charge $n \over m_1$ with strength $m_1$. Vice versa,
electrons of layer 2 see fractionally charged electrons of layer 1
with strength $n \over m_2$. Now, the total flux that each electron
of layer 1 sees is
\beq
\Phi^{\tag{1}} \equiv m_1 ( N^{\tag{1}} - 1 + \frac{n}{m_1} N^{\tag{2}} ) \, ,
\label{charge1}
\eeq
while every electron of layer 2 sees a total of
\beq
\Phi^{\tag{2}} \equiv m_2 ( N^{\tag{2}} - 1 + \frac{n}{m_2} N^{\tag{1}} )
\label{charge2}
\eeq
units of flux. The crucial point is that at the large scale limit, the
two layers are in equilibrium and therefore
$\Phi^{\tag{1}} = \Phi^{\tag{2}}$, which is
\beq
(N^{\tag{1}} - 1 ) m_1 + N^{\tag{2}} n =
N^{\tag{1}} n + (N^{\tag{2}}-1) m_2 \, .
\label{equilibrium}
\eeq
Put in words, regardless of whichever layer the electron is in, the
number of flux each sees in its surrounding is the same. This is not
surprising because both layers are radiated with the same magnetic field.
Using the usual expression of $\nu$ in the study of quantum Hall effect
\beqa
\nu \is \frac{\hbox{\rm total no. of electrons}}{\hbox{\rm total no. of flux}}
\\[4mm]
\is \frac{N^{\tag{1}} + N^{\tag{2}}} {N^{\tag{1}} m_1 + N^{\tag{2}} n}
\, ,
\label{usual}
\eeqa
Taking the limit $N^{\tag{1}} \rightarrow \infty$ and $N^{\tag{2}} \rightarrow
\infty$,
we obtain from (\ref{equilibrium}),
\beq
\nu \rightarrow  \frac{m_1 + m_2 - 2n}{m_1 m_2 - n^2} \, .
\eeq
It is gratifying that the calculation of $\nu$ from the inverse matrix
(\ref{inverse}) as suggested by the effective
topological field theory \cite{genCS}
yields the same result from the usual formula
(\ref{usual}). These two methods are tied together in the
braid group approach.

\subsection{Quasi-hole excitations}
\fpar
In complete analogy to the case of single-layered FQHE, one can consider
quasi-hole excitation on the double-sheeted surface. The quasi-holes too
pick up a tag, corresponding to either of the layers they belong.
Suppose the magnetic field is slightly varied such that there is
a surplus of magnetic flux. Let the number of quasi-holes in the
first layer and the second layer be $N_1$ and $N_2$ respectively.
These quasi-holes are fractionally charged excitations. We denote
their (representation) charges as $c_1$ and $c_2$ respectively.

One can determine the values of $c_1$ and $c_2$ as follows.
As has been discussed in the literature, quasi-holes behave much like
electrons, except they are fractionally charged. Therefore, they also
engage themselves in the puncture phase.
In addition to (\ref{charge1}) that each electron of tag $\tag{1}$
sees, it also senses the flux carried by the quasi-holes:
\beq
m_1 c_1 N_1 + n c_2 N_2 = N_1 \, .
\eeq
Similarly, each electron of tag $\tag{2}$ senses an additional
\beq
m_2 c_2 N_2 + n c_1 N_1 = N_2
\eeq
units of flux. Since both layers are immersed in the same background
magnetic field, small variation in the field strength should result in
the same number of extra fluxes (which are buffered by the quasi-holes).
In other words, $N_1 = N_2$.

Therefore, we get the following two equations:
\beqa
m_1 c_1 + n c_2 \is 1 \, , \\
m_2 c_2 + n c_1 \is 1 \, .
\eeqa
Written in the matrix form, we have
\beq
\left( \begin{array}{cc}
m_1 & n \\
n   & m_2
\end{array} \right)
\left( \begin{array}{c}
c_1 \\
c_2
\end{array} \right)
= \left( \begin{array}{c}
1 \\
1
\end{array} \right) \, .
\label{matrix form}
\eeq
Once again, the matrix of coupling constants appears.
Solving these two equations, one readily finds that
\beqa
c_1 \is \frac{m_2 - n}{\Delta} \, , \\
c_2 \is \frac{m_1 - n}{\Delta} \, .
\eeqa
This result was first obtained by Wilczek \cite{Wilczek-92} who applied the
adiabatic theorem discussed in \cite{ASW}. It is curious to notice that
$c_1$ is equal to the sum of the entries in the first row of the inverse
matrix of the coupling matrix, and $c_2$ the sum of the entries in
the second row. In fact, the matrix equation
(\ref{matrix form}) expresses these equalities
which are also obtainable from the generalized Chern-Simons approach
\cite{genCS}.

With the physical picture offered by the path integral representation of
the braid group,
we can proceed to write down the many-quasi-hole wavefunction of the
bilayered system:
\beqa
\psi & \equiv & \prod_{a < b} ( u_a^{\tag{1}} - u_b^{\tag{1}} )^{m_1 c_1^2}
\prod_{a, b} ( u_a^{\tag{1}} - u_b^{\tag{2}} )^{n c_1 c_2}
\prod_{a < b} ( u_a^{\tag{2}} - u_b^{\tag{2}} )^{m_2 c_2^2} \n
& & \times \,
\prod ( w_a^{\tag{1}} - u_b^{\tag{1}} )^{m_1 c_1}
\prod ( w_a^{\tag{2}} - u_b^{\tag{2}} )^{m_2 c_2}
\prod ( w_a^{\tag{1}} - u_b^{\tag{2}} )^{n c_2}
\prod ( w_a^{\tag{2}} - u_b^{\tag{1}} )^{n c_1} \n
& & \times \,
\exp ( - \frac{1}{4 \ell^2 c_1} \sum_a^{N_1} |u_a^{\tag{1}}|^2
- \frac{1}{4 \ell^2 c_2} \sum_a^{N_2} |u_a^{\tag{2}}|^2 )
\, \, \psi_{m_1, m_2, n}
\label{qh-bilayer}
\eeqa
This wavefunction is an exact solution of the bilayered version of
the ground state equations:
\beq
\left( {\cal D}_a^{\tag{i}} + \frac{ \frac{e}{c^{\tag{i}}} B}{ 4 \hbar c}
{\ol u}_a^{\tag{i}} \right) \, \psi  = 0 \, ,
\label{D-gse}
\eeq
where
\beqa
{\cal D}_a^{\tag{1}} \!\! & \equiv & \!\!
\partial_{u^{\tag{1}}_a}
- m_1 \sum_{b \neq a}^{N_1} \frac{ c_1 \times c_1 }
{ u_a^{\tag{1}} - u_b^{\tag{1}} }
-n \sum_{b}^{N_2} \frac{c_1 \times c_2}
{ u_a^{\tag{1}} - u_b^{\tag{2}} }
- m_1 \sum_{b}^{N^{\tag{1}}} \frac{c_1 \times 1}
{ u_a^{\tag{1}} - w_b^{\tag{1}} }
- n \sum_{b}^{N^{\tag{2}}} \frac{c_1 \times 1}
{ u_a^{\tag{1}} - w_b^{\tag{2}} }
\n
\is \partial_{u^{\tag{1}}_a}
- m_1 c_1 \sum_{b \neq a}^{N_1} \frac{ 1 \times c_1 }
{ u_a^{\tag{1}} - u_b^{\tag{1}} }
- m_1 c_1  \sum_{b}^{N_2} \frac{1 \times \frac{n}{m_1} c_2}
{ u_a^{\tag{1}} - u_b^{\tag{2}} }
\n
\! \! & & \! \! - \, \,
m_1 c_1 \sum_{b}^{N^{\tag{1}}} \frac{ 1 \times 1}
{ u_a^{\tag{1}} - w_b^{\tag{1}} }
- m_1 c_1 \sum_{b}^{N^{\tag{2}}} \frac{ 1 \times \frac{n}{m_1}}
{ u_a^{\tag{1}} - w_b^{\tag{2}} }
\, ,
\eeqa
and
\beqa
{\cal D}_a^{\tag{2}} \!\! & \equiv & \!\!
\partial_{u^{\tag{2}}_a}
- m_2 \sum_{b \neq a}^{N_2} \frac{ c_2 \times c_2 }
{ u_a^{\tag{2}} - u_b^{\tag{2}} }
-n \sum_{b}^{N_1} \frac{c_2 \times c_1}
{ u_a^{\tag{2}} - u_b^{\tag{1}} }
- m_2 \sum_{b}^{N^{\tag{2}}} \frac{c_2 \times 1}
{ u_a^{\tag{2}} - w_b^{\tag{2}} }
- n \sum_{b}^{N^{\tag{1}}} \frac{c_2 \times 1}
{ u_a^{\tag{2}} - w_b^{\tag{1}} }
\n
\is \partial_{u^{\tag{2}}_a}
- m_2 c_2 \sum_{b \neq a}^{N_2} \frac{ 1 \times c_2 }
{ u_a^{\tag{2}} - u_b^{\tag{2}} }
- m_2 c_2  \sum_{b}^{N_1} \frac{1 \times \frac{n}{m_2} c_1}
{ u_a^{\tag{2}} - u_b^{\tag{1}} }
\n
\! \! & & \! \! - \, \,
m_2 c_2 \sum_{b}^{N^{\tag{2}}} \frac{ 1 \times 1}
{ u_a^{\tag{2}} - w_b^{\tag{2}} }
- m_2 c_2 \sum_{b}^{N^{\tag{1}}} \frac{ 1 \times \frac{n}{m_2}}
{ u_a^{\tag{2}} - w_b^{\tag{1}} }
\, .
\eeqa

The statistical parameters of the
quasi-holes are easily readable from (\ref{qh-bilayer}) and they are
\beqa
\frac{\theta^{\tag{11}}}{\pi} \is m_1 c_1^2 \, , \\
\frac{\theta^{\tag{12}}}{\pi} \is n c_1 c_2 \, , \\
\frac{\theta^{\tag{11}}}{\pi} \is m_2 c_2^2 \, .
\eeqa
We have checked that these values coincide with those calculated
with the
generalized Chern-Simons approach \cite{genCS}.
$\frac{\theta^{\tag{12}}}{\pi}$ is the {\em mutual statistics}
of the quasi-holes residing in different layers.

\section{Discussions}
\subsection{Inter-layer correlation}
\fpar
The braid group formalism leading to the ground state equations (\ref{D-gse})
does not specify the relation between $m_1, m_2$ and $n$.
Much like $m_1$ and $m_2$ are restricted to odd integers due to the
Pauli principle, we shall explore the possible values of $n$ from
physical considerations.

First of all, $n$ must be an integer because the charge carriers
are electrons and the wavefunction of the total system must be single-valued
with respect to monodromy.
It is also
evident that if $\delta = \frac{d}{\ell}$ is large, which corresponds
to two isolated single layers, the inter-layer correlation should weaken, if
not vanishes. On the other extreme, when $d$ is zero, which is the
usual single-layered setup, the total number of (polarized)
electrons in the layer is $N^{\tag{1}} + N^{\tag{2}}$ and $m_1 = m_2 = n$.
In this instance, the $2 \times 2$ matrix (\ref{k-mat})
collapses to a number $m_1$ (otherwise it becomes singular).
The point is that $n$ strictly depends on $\delta$, which is an
indicator of the onset of inter-layer correlation.

Next, we want to propose that $n$ should be an odd number as well.
The physical picture emerging from electrons moving on a double-sheeted
surface is that electrons with tag ${\tag{1}}$ and those with
${\tag{2}}$ should be {\em effectively}
indistinguishable as well. This is because the
magneto-transport measurement is performed on the
total system and thus cannot possibly distinguish which of the
two layers the electrons are from. In a certain sense, the bilayered system
is analogous to a
d. c. circuit of two resistors in parallel. Before and after the
electrons flow through the resistors, there is no way to tell which of the
two resistors each is going to pass or has passed.
Assigning a tag to the coordinates does not make the electrons distinguishable,
just like giving a name $a$ does not make the electrons of the same layer
distinguishable. Therefore, invoking Pauli principle, $n$ is
restricted to be an odd number.

Though it is far more difficult to justify, we feel that $n$ should be smaller
than $m_1$ and $m_2$. The braid group approach reveals that these numbers
are the strengths with which an electron sees its counterparts as
punctures. Given this interpretation, it is therefore expected that
electrons sense those of the same layer with greater strength than those
from different layer. Summarizing what have been said so far,
$n = n ( \delta )$, and it monotonously decreases from $m_1$ ( or $m_2$,
whichever is smaller ) to 0.

\subsection{Mutual statistics}
\fpar
It is now clear that mutual statistics arises in bilayered system
because of $k^{\tag{12}} \neq 0 $.
Since $k^{\tag{12}} = n$ is a function of $\delta$ as
discussed earlier, it is the consequence of inter-layer
correlation. In the braid group approach, electrons from
both layers participate in the puncture phase on the double-sheeted
surface. Within an appropriate range of $\delta$, $n$ takes on non-trivial
value and mutual statistics is realized.

\subsection{Coupling matrix and multi-layered FQHE}
\fpar
The method we use to study the bilayered FQHE can be easily extended to
multi-layered systems. It is straightforward to
generalize the construction discussed
in section 3 by replacing double-sheeted surface with multi-sheeted plane.
The path integral representation of the braid group now involves a
$M \times M$ matrix of coupling contants:
\beq
\left( \begin{array}{llll}
k^{\tag{11}} & k^{\tag{12}} & \cdots & k^{\tag{1M}} \\
k^{\tag{21}} & k^{\tag{22}} & \cdots & k^{\tag{2M}} \\
\vdots       & \vdots       & \ddots & \vdots       \\
k^{\tag{M1}} & k^{\tag{M2}} & \cdots & k^{\tag{MM}}
\end{array} \right)
\label{coupling matrix}
\eeq
The coupling matrix (\ref{coupling matrix}) compactly encodes the topological
information of an ensemble of free electrons moving in a multi-sheeted surface.
The filling factor of the multi-layered Hall fluid is the sum of all the
entries of the inverse matrix of (\ref{coupling matrix}).
The dimension of the coupling matrix corresponds to the number of layers of the
electronic system.
When the Hall media is single-layered, the coupling matrix is $1 \times 1$ and
(\ref{coupling matrix}) reduces to just one number, the inverse of which
gives the filling factor of Laughlin type.
Therefore, the coupling matrix is a generalization of the usual
description of single-layered FQHE.
For example, consider the following coupling matrix:
\beq
\left( \begin{array}{ccc}
m_1 & n   & 0  \\
n   & m_2 & n  \\
0   & n   & m_3
\end{array} \right) \, .
\label{3 layers}
\eeq
It describes a Hall media of three layers. Calculating the sum of all the
entries of the inverse matrix of (\ref{3 layers}), we find that the filling
factor is
\beq
\nu = \frac{m_1 m_2 + m_1 m_3 + m_2 m_3 - 2 m_1 n - 2 m_3 n}
{m_1 m_2 m_3 - m_1 n^2 - m_3 n^2} \, .
\label{3 nu}
\eeq
One can check that (\ref{3 nu}) coincides with the usual calculation.
The equilibrium condition is
\beq
m_1 [ ( N^{\tag{1}} - 1 ) + N^{\tag{2}} \frac{n}{m_1} ]
= m_2 [ N^{\tag{1}} \frac{n}{m_2} +
       ( N^{\tag{2}} - 1 ) + N^{\tag{3}} \frac{n}{m_2} ]
= m_3 [ N^{\tag{2}} \frac{n}{m_3} + ( N^{\tag{3}} - 1 ) ] \, .
\eeq
In the limit $N^{\tag{i}} \rightarrow \infty $ for all $i$, the formula
(\ref{usual}) yields the same value as (\ref{3 nu}). If $m_i = 3$ and $n = 1$,
$\nu = \frac{5}{7}$. This should
be the 3-layered analogue of the $\nu = \half$
effect observed by \cite{Eisenstein}.

\subsection{3-dimensional FQHE}
\fpar
Suppose there are many layers, such that the dimension $M$ of the
coupling matrix is large. It is then possible to contemplate a multi-layered
FQHE characterized by the following coupling matrix:
\beq
\left( \begin{array}{cccccc}
m_1    &  n     &  0     & \cdots  &  \cdots   &  0      \\
n      &  m_2   &  n     & \cdots  &  \cdots   &  \vdots \\
0      &  n     & m_3    & \ddots  &           &  \vdots \\
\vdots & \vdots & \ddots & \ddots  &  \ddots   &  \vdots \\
\vdots & \vdots &        & \ddots  &  m_{M-1}  &  n      \\
0      &  0     & \cdots & \cdots  &  n        &  m_M
\end{array} \right)
\label{3-d}
\eeq
The symmetric tridiagonal coupling matrix (\ref{3-d})
is an interesting case because it corresponds to the
nearest-neighour inter-layer correlation which is more likely to
be realized experimentally. Now when the number of layers $M$
is large, the electronic system is literally 3-dimensional!
Since each layer correlates only with its nearest layers,  the model
can be regarded as locally 2-dimensional. By juxtaposing narrow
quantum wells alternately with barriers of appropriate width,
one may create 3-dimensional FQHE in the laboratory.

\section{Concluding remarks}
\fpar
The representation theory of the braid group over the multi-sheeted surface
has clearly given us a useful basis to consider the physics of
multi-layered Hall media. The most interesting feature of the multi-layered
FQHE, as compared to single-layered FQHE, is the notion of
fractional mutual statistics. We have proposed a Hamiltonian
(\ref{bilayer H}) for which the
YMG ans\"atz (\ref{YMG}) is the {\em exact} solution of the ground state
equations. As in the single-layered case, the topological order of
the multi-layered system lies in the ``puncture phase" wherein each particle
sees the rest as punctures on the multi-sheeted surface.

The braid group approach presents us a tool, namely the coupling matrix
(\ref{coupling matrix}) to capture the filling factors, fractional
charges and statistics of the quasi-excitations of the multi-layered FQHE.
It is worthwhile to highlight that a $\nu = \frac{5}{7}$ effect is
predicted for a sample with three layers of electron fluid. Finally,
if one has the technology to grow many layers of heterojunctions, it is
even possible to realize a 3-dimensional FQHE described by the
coupling matrix (\ref{3-d}).

\subsection*{Acknowledgement}
I would like to thank C. H. Lai for encouragement and discussions,
and C. L. Ho for a useful conversation on ``duality".

\subsection*{Note added}
After this paper was submitted for publication, I learned
that Ezawa and Iwazaki had
also independently considered the concept of mutual statistics
(which they called relative statistics) in \cite{EI}.


\begin{thebibliography}{99}

\bibitem{Wilczek}
For review and references to earlier original works,
see {\em Fractional Statistics and Anyon Superconductivity},
ed. F. Wilczek, (World Scientific, Singapore, 1990).

\bibitem{long}
R. G. Clark, J. R. Mallett, S. R. Haynes, J. J. Harris and C. T. Foxon,
Phys. Rev. Lett. {\bf 60} (1988) 1747; J. A. Simmons, H.-P. Wei, L. W. Engel,
D. C. Tsui and M. Shayegan, Phys. Rev. Lett. {\bf 63} (1989) 1731.

\bibitem{Chang-Cunningham}
A. M. Chang and J. E. Cunningham, Solid State Commun. {\bf 72} (1989) 651.

\bibitem{Laughlin}
R. B. Laughlin, Phys. Rev. Lett. {\bf 50} (1983) 1395.
See also his lecture in \cite{Wilczek}  pp. 267-268.

\bibitem{superconductivity}
R. B. Laughlin, Science {\bf 242} (1988) 525; Phys. Rev. Lett. {\bf 60}
(1988) 2677; A. Fetter, C. Hanna and R. B. Laughlin,
Phys. Rev. {\bf B39} (1989) 9676;
Y.-H. Chen, F. Wilczek, E. Witten and B. I. Halperin,
Int. J. Mod. Phys. {\bf B3} (1989) 1001;

\bibitem{supcond}
For later development,
see the relevant papers in the {\em Proceedings of the TCSUH Workshop on
Physics and Mathematics of Anyons} ed. S. S. Chern, C. W. Chu and C. S. Ting,
Int. J. Mod. Phys. {\bf B5} (1991) 1485.

\bibitem{nonabelian}
E. Verlinde, ``A note on braid statistics and the non-abelian
Aharonov-Bohm effect", preprint IASSNS-HEP-90/60, Jan 1991.

\bibitem{Lai-Ting}
C. H. Lai and C. Ting, Phys. Lett. {\bf B265} (1991) 341-346.

\bibitem{higher genus}
T. Einarsson, Phys. Rev. Lett. {\bf 64} (1990) 1995;
Y. S. Wu, in \cite{supcond} 1649.

\bibitem{Wilczek-92}
F. Wilczek, ``Disassembling Anyons", preprint IASSNS-HEP-91/70, Jan 1992.

\bibitem{genCS}
A. Zee, in \cite{supcond} 1629;
J. Fr\"ohlich and A. Zee, Nucl. Phys. {\bf B364} (1991) 517.

\bibitem{Suen}
Y. W. Suen, L. W. Engel, M. B. Santos, M. Shayegan and D. C. Tsui,
Phys. Rev. Lett. {\bf 68} (1992) 1379.

\bibitem{Eisenstein}
J. P. Eisenstein, G. S. Boebinger, L. N. Pfeiffer, K. W. West and
Song He, Phys. Rev. Lett. {\bf 68} 1383.

\bibitem{GWW}
M. Greiter, X. G. Wen and F. Wilczek, ``Paired Hall States in Double
Layer Electron Systems", preprint IASSNS-HEP-92/1, Jan 1992.

\bibitem{YMG}
D. Yoshioka, A. MacDonald and S. Girvin, Phys. Rev. {\bf B39} (1988) 1932.

\bibitem{Ting-Lai-1}
C. Ting and C. H. Lai, Mod. Phys. Lett. {\bf B5} (1991) 1293-1299.

\bibitem{Ting-Lai-2}
C. Ting and C. H. Lai, ``Spinning Braid Group Representation and
the Fractional Quantum Hall Effect", preprint NUS/HEP/92011,
(Jan, 1992) (hepth@xxx/~9202024).

\bibitem{Wu}
Y. S. Wu, Phys. Rev. Lett. {\bf 52} (1984) 2103-2106.

\bibitem{KZ}
V. Knizhnik and A. Zamolodchikov, Nucl. Phys. {\bf B247} (1984) 83.

\bibitem{Ahlfors}
For a lucid discussion, see
L. V. Ahlfors, {\em Complex Analysis}, third edition, (McGraw-Hill, 1979).

\bibitem{Frohlich-Kerler}
J. Fr\"ohlich and T. Kerler,
Nucl. Phys. {\bf B354} (1991) 369.

\bibitem{Halperin}
B. I. Halperin, Helv. Phys. Acta {\bf 56} (1983) 75.

\bibitem{ASW}
D. Arovas, J. R. Schrieffer and F. Wilczek, Phys. Rev. Lett. {\bf 53}
(1984) 722.

\bibitem{EI}
Z. F. Ezawa and A. Iwazaki, ``Chern-Simons Gauge Theory for Double-Layer
Electron System", preprint TU-402, Nisho-18 (May, 1992).
\end{thebibliography}
\end{document}